\documentstyle[11pt,epsfig]{article}
\normalbaselines
\newcounter{eqnletter}[equation]
\catcode`\@=11
\@addtoreset{equation}{section}   
\catcode`\@=12

\begin{document}

\begin{center}

{\LARGE\bf On the Duality of \\[0.5cm]
Quasi-Exactly Solvable Problems}

\vskip 1cm

{\large {\bf Agnieszka Krajewska\footnote{akraj@mvii.uni.lodz.pl}, 
             Alexander Ushveridze\footnote{alexush@mvii.uni.lodz.pl}}}\\
{\bf and}\\ 
{\large {\bf Zbigniew Walczak\footnote{walczak@mvii.uni.lodz.pl} }}

\vskip 0.1 cm

Department of Theoretical Physics, University of Lodz,\\
Pomorska 149/153, 90-236 Lodz, Poland

\end{center}
\vspace{1 cm}
\begin{abstract}

It is demonstrated that quasi-exactly solvable models of
quantum mechanics admit an interesting duality transformation
which changes the form of their potentials and inverts the
sign of all the exactly calculable energy levels. This 
transformation helps one to reveal some new features
of quasi-exactly solvable models and associated
orthogonal polynomials.

\end{abstract}

\newpage

\section{Introduction}

In this paper we demonstrate that quasi-exactly solvable models 
of quantum mechanics (i.e. models admitting exact solutions
only for certain limited parts of the spectrum)\footnote{
For details see original papers \cite{zasl, turush, ush, tur} and
reviews \cite{shif, ushbook}.} normally appear 
in the form of dublets. The models forming a given dublet  look 
differently but their exactly calculable energy levels exactly 
coincide up to a sign transformation. We call such models {\it dual} 
to each other. Note that the duality property does not hold for 
other (exactly noncalculable) levels. 

In this paper, we consider simplest quasi-exactly solvable 
models introduced in ref. \cite{turush} and 
represented by even polynomials of degree six (section 2). 
We construct the dual pairs of these models (section 3) and
show that the duality property enables one to estimate the intervals 
of the spectra occupied by exactly calculable energy levels (section 4).
We also demonstrate that the duality property can be used 
for finding classical counterparts of the phenomenon of quasi-exact 
solvability (section 5) and for establishing new remarkable properties of
non-standard orthogonal polynomials associated with
quasi-exactly solvable models (section 6).
A brief discussion of the general case will be given in the
concluding section 7.

\section{The model}

The Schr\"odinger equation for the simplest sequence of quasi-exactly
solvable models  reads
\begin{eqnarray}
\left\{-\hbar^2\frac{\partial^2}{\partial x^2}+ 
x^2(ax^2+b)^2-\hbar a (2M+3)x^2\right\}
\Psi(x)=E\Psi(x),
\label{2.1}
\end{eqnarray}
where $b$ is real, $a>0$ is a positive real number and $M=0,1,\ldots$
is an arbitrary non-negative integer. 

The explicit solutions of this equation can be represented in the form
\begin{eqnarray}
E=\hbar\left[(2M+1)b+2a\sum_{i=1}^M \xi_i^2\right]
\label{2.2}
\end{eqnarray}
and
\begin{eqnarray}
\Psi(x)=\prod_{i=1}^M (x-\xi_i)
\exp\left[-\frac{1}{\hbar}\left(\frac{bx^2}{2}+
\frac{ax^4}{4}\right)\right],
\label{2.3}
\end{eqnarray}
where the numbers $\xi_i, \ i=1,\ldots, M$ (playing the
role of wavefunction zeros) satisfy the system of numerical equations
\begin{eqnarray}
{\sum_{k=1}^M}' \frac{\hbar}{\xi_i-\xi_k}=b\xi_i+a\xi_i^3,
\quad i=1,\ldots, M
\label{2.4}
\end{eqnarray}
and the following additional condition
\begin{eqnarray}
\sum_{i=1}^M \xi_i=0.
\label{2.5}
\end{eqnarray}
The proof of formulas (\ref{2.2}) -- (\ref{2.5}) is trivial.
For details see the proof of analogous formulas in ref. \cite{ushbook}.
The number of exactly calculable energy levels and
wavefunctions is, obviously, given by the number of
solutions of the system (\ref{2.4}) -- (\ref{2.5}). At
first sight, this system is overdetermined, because it
contains $M+1$ equations for only $M$ unknowns $\xi_i, \ i=1,\ldots,M$.
However, it can be shown that, due to the symmetry of this system 
under reflections $\xi_i\rightarrow -\xi_i$, it still has non-trivial
solutions for any $M$, and their number is equal to
$[M/2]+1$, where $[\ ]$ is a standard notation for an
integer part. Note that the parity of the obtained energy
levels coincides with the parity of $M$. It can also be shown
that the numbers $\xi_i$ satisfying the system (\ref{2.4})
-- (\ref{2.5}) are either real or purely imaginary.

\section{Duality and self-duality}

It is easily seen that formulas (\ref{2.1}) -- (\ref{2.5}) for the ``plus''
model (with $b=|b|$) can be reduced to the same formulas
for ``minus'' model (with $b=-|b|$) after the changes
\begin{eqnarray}
x\rightarrow ix, \qquad \xi_i\rightarrow i\xi_i, \qquad
E\rightarrow -E. 
\label{3.1}
\end{eqnarray}
Hereafter we shall call the models (\ref{2.1}),
differing from each other by the sign of the parameter $b$,
{\it dual} to each other and the transformation (\ref{3.1})
--- the {\it duality transformation}. We also shall use the notations
\begin{eqnarray}
V(x)=x^2(ax^2+b)^2-\hbar a (2M+3)x^2,
\label{3.2}
\end{eqnarray}
for the potential of the original model (\ref{2.1}), and
\begin{eqnarray}
\bar V(x)=x^2(ax^2-b)^2-\hbar a (2M+3)x^2,
\label{3.3}
\end{eqnarray}
for its dual. If the parameter $b$ is zero
then the potentials $V(x)$ and $\bar V(x)$ exactly coincide
and read
\begin{eqnarray}
V(x)=\bar V(x)=a^2x^6-\hbar a (2M+3)x^2.
\label{3.4}
\end{eqnarray}
It is natural to call the model (\ref{3.4}) {\it self-dual}.

\section{Spectral properties}

From formula (\ref{3.1}) it follows that, if $E$ is an
exactly calculable energy level in model (\ref{3.2}), then
$\bar E=-E$ is also a certain exactly calculable energy
level in the dual model (\ref{3.3}). It is interesting to
establish a relationship between the quantum numbers
(ordinal numbers) of these levels. Assume that the energy level $E$ describes
the $K$th excitation, $E=E_K$. Then, according to the oscillation theorem,
the wavefunction (\ref{2.3}), $\Psi(x)=\Psi_K(x)$, should have $K$ real zeros
(nodes) and, therefore, $K$ numbers $\xi_i$ should lie on the real $x$-axis. 
In this case, the $M-K$ remaining numbers 
$\xi_i$ should lie on the imaginary $x$-axis. But from
formula (\ref{3.1}) it follows that, after the duality transformation,
all the imaginary numbers $\xi_i$ describing the imaginary wavefunction zeros
in the model (\ref{3.2}) become real, $\bar\xi_i=i\xi_i$,
and take the meaning of real wavefunction zeros (nodes) in
the dual model (\ref{3.3}). But this means that 
$\bar \Psi(x)=\bar\Psi_{M-K}(x)$ the corresponding
energy level describes the $(M-K)$th excitation, $\bar
E=\bar E_{M-K}$. This leads us to the simple relations
\begin{eqnarray}
E_K=-\bar E_{M-K}, \qquad \Psi_K(x)\sim\bar\Psi_{M-K}(ix).
\label{4.1}
\end{eqnarray}
In particular, the minimal exactly calculable energy level in  a given
quasi-exactly solvable model (\ref{3.2})  
coincides with the maximal exactly calculable energy level in
the corresponding dual model (\ref{3.3}) taken with sign minus!
For the self-dual model (\ref{3.4}) the situation becomes
even more transparent. In this case, because of the
coincidence of the sets $\{E\}$ and $\{\bar E\}$, we obtain
\begin{eqnarray}
E_K=-E_{M-K}, \qquad \Psi_K(x)\sim\Psi_{M-K}(ix).
\label{4.2}
\end{eqnarray}

The abovementioned facts enable one to detemine the intervals of the spectra
of models (\ref{3.2}), (\ref{3.3}) and (\ref{3.4})
occupied by the exactly calculable levels. We denote these intervals
by $[E_{min},E_{max}]$ and $[\bar E_{min},\bar E_{max}]$, respectively.
The duality property means that 
\begin{eqnarray}
E_{max}=-\bar E_{min}, \qquad \bar E_{max}=-E_{min}.
\label{4.3}
\end{eqnarray} 
On the other hand, we have the obvious inequalities
\begin{eqnarray}
E_{min}>\mbox{min}\ V(x)  , \qquad \bar E_{min}>\mbox{min}\
\bar V(x).
\label{4.4}
\end{eqnarray} 
Comparing (\ref{4.3}) with (\ref{4.4}), we obtain
\begin{eqnarray}
E_{max}<-\mbox{min}\ \bar V(x)  , \qquad \bar E_{max}<-\mbox{min}\ V(x).
\label{4.5}
\end{eqnarray} 
This means that
\begin{eqnarray}
[E_{min},E_{max}]\subset [\mbox{min}\ V(x),-\mbox{min}\
\bar V(x)],\quad
[\bar E_{min},\bar E_{max}]\subset [\mbox{min}\ \bar V(x),-\mbox{min}\
V(x)],
\label{4.6}
\end{eqnarray} 
where
\begin{eqnarray}
\mbox{min}\ V(x) =
\left\{
\begin{array}{l}
0, \quad b>\sqrt{ac}, \\
2(27a)^{-1}\left(-2b+\sqrt{b^2+3ca}\right)
\left(-3ca+b^2+b\sqrt{b^2+3ca}\right), \quad b<\sqrt{ac}
\end{array}
\right.
\label{4.7}
\end{eqnarray} 
and
\begin{eqnarray}
\mbox{min}\ \bar V(x) =
\left\{
\begin{array}{l}
0, \quad -b>\sqrt{ac}, \\
2(27a)^{-1}\left(+2b+\sqrt{b^2+3ca}\right)
\left(-3ca+b^2-b\sqrt{b^2+3ca}\right), \quad -b<\sqrt{ac}
\end{array}
\right.
\label{4.8}
\end{eqnarray} 
Here we introduced a new quantity
\begin{eqnarray}
c=(2M+3)\hbar.
\label{4.9}
\end{eqnarray}
For the self-dual model (\ref{3.4}), formulas
(\ref{4.6}) -- (\ref{4.8}) become simpler. Taking in them
$b=0$, we obtain
\begin{eqnarray}
[E_{min},E_{max}]=[\bar E_{min},\bar E_{max}]
\subset \left[-\frac{\sqrt{12a}c^{3/2}}{9},
\frac{\sqrt{12a}c^{3/2}}{9} \right].
\label{4.10}
\end{eqnarray} 
The examples of the dual and self-dual potentials with the
intervals occupied by exactly calculable energy levels are
depicted in figures 1 and 2.

\begin{figure}[h]
\begin{center}
\epsfxsize\textwidth 
\epsffile{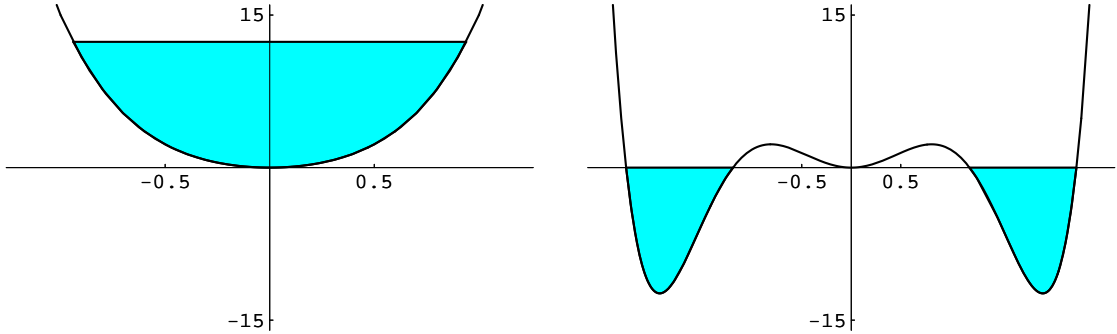}
\end{center}
\caption[]{The form of the potential $V(x)=x^2(ax^2+b)^2-acx^2$
(left picture) and its dual $\bar V(x)=
x^2(ax^2-b)^2-acx^2$ (right picture) for $a=1$,  $b=3.3$ and $c=3.5$.
The gray domains correspond to the parts of spectra occupied by
the exactly calculable levels. Here $[E_{min},E_{max}]=[0,12.37]$
and $[\bar E_{min},\bar E_{max}]=[-12.37,0]$. }
\end{figure}
\begin{figure}[h]
\begin{center}
\epsfxsize=\textwidth 
\epsffile{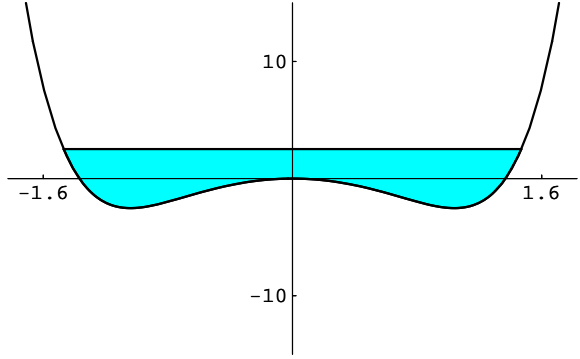}
\end{center}
\caption[]{The form of the self-dual potential $V(x)=\bar V(x)=
a^2x^6-acx^2$ for $a=1$ and $c=3.5$. The gray domain corresponds to the
part of the spectrum occupied by exactly calculable levels. Here
$[E_{min},E_{max}]=[-2.52,+2.52]$.}
\end{figure}


\section{The classical limit}

It is naturally to ask what is the classical analogue of
the phenomenon of quasi-exact solvability of models (\ref{2.1})?
As far as we know, this question has never been raised in
the literature. To answer it, we should understand what is the
difference between the solutions of the classical 
equation of motion corresponding to the cases with $E\in
[E_{min},E_{max}]$ and $E\not\in [E_{min},E_{max}]$. For
this it is very helpful to use the duality property.

First note that for small $\hbar$ the spacing between the
neighbouring exactly calculable energy levels is of order
of $\hbar$. If $\hbar\rightarrow0$ then the lowest exactly
calculable levels in models (\ref{3.2}) and (\ref{3.3}) 
approach the absolute minima of potentials $V(x)$ and $\bar
V(x)$. For this reason, the classical limit of the relation
(\ref{4.6}) reads
\begin{eqnarray}
[E_{min},E_{max}]=[\mbox{min}\ V_0(x),-\mbox{min}\
\bar V_0(x)],\quad
[\bar E_{min},\bar E_{max}]=[\mbox{min}\ \bar V_0(x),-\mbox{min}\
V_0(x)],
\label{5.1}
\end{eqnarray} 
where by $V_0(x)$ and $\bar V_0(x)$ we denoted the classical
limits of the potentials $V(x)$ and $\bar V(x)$.

For determining the classical potentials remember that the
original ones contain two arbitrary parameters $\hbar$ and $M$.
Therefore, the result of classical limit will depend of how
fast $M$ tends to infinity if $\hbar$ tends to zero.
This forces one to consider separaely two cases: 1) $\hbar\rightarrow0$
and $c\rightarrow0$ and 2) $\hbar\rightarrow0$ and
$c\neq 0$. These limits we shall call respectively the {\it
soft} and {\it hard} classical limits.

{\bf The soft classical limit.}
In the soft classical limit the potentials (\ref{3.2})
and (\ref{3.3}) take the form
\begin{eqnarray}
V_0(x)=x^2(ax^2+b)^2,\qquad \bar V_0(x)=x^2(ax^2-b)^2.
\label{5.2}
\end{eqnarray}
In this case we have
\begin{eqnarray}
\mbox{min}\ V_0(x)=0,\qquad \mbox{min}\ \bar V_0(x)=0,
\label{5.2a}
\end{eqnarray}
and therefore, the parts of the spectrum occupied
by the exactly calculable levels are collapsing into a single
(ground state) energy level $E=\bar E=0$. For this reason,
it is naturally to expect that 
the classical analogue of the phenomenon of quasi-exact solvability
should lie in a comparative simplicity of the ground
state solution ($E=\bar E=0$) for the model (\ref{3.2}), (\ref{3.3}) 
with respect to all other solutions ($E>0$).

In order to make sure that this is really so, consider, for
example, the expression for the subbarier classical
action $S(x)$ in the model (\ref{3.2}). Remember, that function
$S(x)$ is a natural classical counterpart of the quantum wavefunction
$\Psi(x)$ and is related to it by the formula
$\Psi(x)\approx e^{-S(x)/\hbar}$. Solving the corresponding
Hamilton -- Jacobi equation we obtain
\begin{eqnarray}
S(x)=\int dx \sqrt{
x^2(a x^2 + b)^2 - E}, \qquad E\ge 0.
\label{5.3}
\end{eqnarray}
In some sense this is an explicit expression which, after
evaluating the integral, can be reduced to the combination of
tabulated elliptic integrals. Let us however 
consider the particular case with $E=0$ when some
analogue of quasi-exact solvability is expected.
In this case the integral (\ref{5.3}) can be evaluated
explicitly and the result takes a very elegant and simple form
\begin{eqnarray}
S(x)=\frac{a x^4}{4}+ \frac{b x^2}{2}, \qquad E=0.
\label{5.4}
\end{eqnarray}
We see that this case is distinguished from all other cases
by the fact that the action function is
in this case regular, without any branch - points and cuts!
Comparing the special solution (\ref{5.4}) with the general
one (\ref{5.3}), we can say with confidence that
the former has all rights to be called "exact solution". 

{\bf The hard classical limit.}
In the hard classical limit the potentials of the models (\ref{3.2})
and (\ref{3.3}) take the form
\begin{eqnarray}
V_0(x)=x^2(ax^2+b)^2-acx^2,\qquad \bar V_0(x)=x^2(ax^2-b)^2-acx^2.
\label{5.5}
\end{eqnarray}
The number of exactly calculable energy levels tends to infinity 
and the spacing between them tends to zero. The sets
of former quantum exactly calculable energy levels
transform into the segments of a continuous
spectrum given by the formula (\ref{5.1}).
The subbarier classical action now reads
\begin{eqnarray}
S(x)=\int dx \sqrt{
x^2(a x^2 + b)^2 -acx^2- E}.
\label{5.6}
\end{eqnarray} 
It is not difficult to see
that, in the complex $x$-plane, the function $S(x)$ has six
square-root-type branch point singularities corresponding to the roots of
the sixth-order algebraic equation $x^2(ax^2+b)^2-cax^2=E$.
These singularities are connected pairwise by cuts. A
simple analysis shows that if
$E\in [E_{min},E_{max}]$, then all the six singularities lie
on the real and imaginary $x$-axes. However, if $E\not\in [E_{min},E_{max}]$, 
then there are singularities lying outside the axes.
The location of these singularities and the corresponding
cuts are depicted in the figure 3.

{\footnotesize
\begin{figure}[h]
\begin{center}
\epsfxsize=\textwidth 
\epsffile{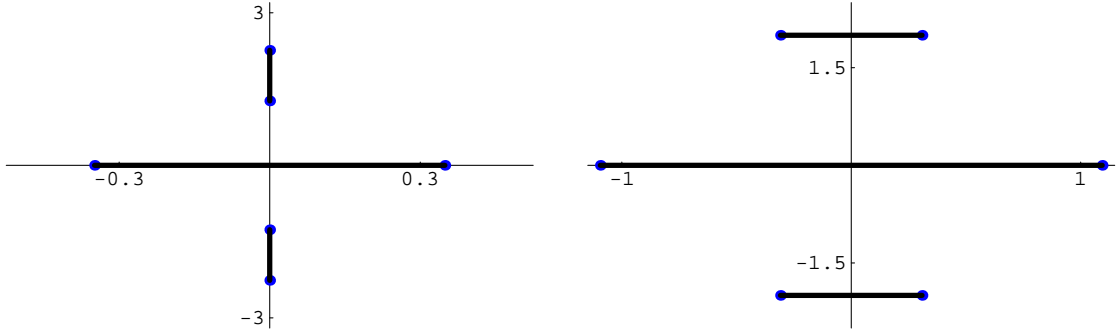}
\end{center}
\caption[]{The complex $x$-plane for the classical
action $S(x)$ (formula (\ref{5.6})). Here, as before,
$a=1$,  $b=3.3$, $c=3.5$ and
$[E_{min}, E_{max}]=[0, 12.37]$ is an interval occupied
by the former exactly
calculable energy levels. In the left picture $E=1\in [E_{min}, E_{max}]$,
while in the right picture $E=20\not\in [E_{min},E_{max}]$.
We see, that the disctribution of
singularities and cuts for these two cases is different. }
\end{figure}
}


So, we see that there is some qualitative difference
between the solutions of classical equations corresponding
to the former exactly calculable and exactly non-calculable
energy levels. At the same time, there is no difference between
the complexity of these solutions. This is quite obvious, because
if $M\rightarrow \infty$ the algebraic equations
determining the exactly calculable states in quantum
quasi-exactly solvable models become infinitely complicated and the difference
between them and any other states dissapears.

\section{The duality and orthogonal polynomials}

The wavefunctions corresponding to exactly calculable
energy levels in models (\ref{3.2}) and (\ref{3.3}) can be
represented in the form
\begin{eqnarray}
\Psi_m(x)&=&x^p P_m\left(\frac{x^2}{\hbar}\right)
\exp\left[-\frac{1}{\hbar}\left(\frac{ax^4}{4}+
\frac{bx^2}{2}\right)\right],\nonumber\\
\bar\Psi_m(x)&=&x^p \bar P_m\left(\frac{x^2}{\hbar}\right)
\exp\left[-\frac{1}{\hbar}\left(\frac{ax^4}{4}-
\frac{bx^2}{2}\right)\right],
\label{6.1}
\end{eqnarray}
where $p=0,1$ is the parity of the solution, and $P_m(t)$, $m=0,1,\ldots,[M/2]$
are certain polynomials of degree $[M/2]$.
From the condition of the orthogonality of wavefunctions
\begin{eqnarray}
\int_{-\infty}^{+\infty}\Psi_m(x)\Psi_n(x)dx=0,\quad
\int_{-\infty}^{+\infty}\bar\Psi_m(x)\bar\Psi_n(x)dx=0,\quad
m\neq n
\label{6.2}
\end{eqnarray}
it follows the orthogonality of polynomials $P_m(t)$ and
$\bar P_m(t)$:
\begin{eqnarray}
\int_{0}^{\infty}P_m(t)P_n(t)\omega(t)dt=0,\quad
\int_{0}^{\infty}\bar P_m(t)\bar P_n(t)\bar\omega(t)dt=0,\quad
m\neq n
\label{6.3}
\end{eqnarray}
with the weight functions
\begin{eqnarray}
\omega(t)=t^{p-\frac{1}{2}}e^{-\frac{at^2}{2}-bt},\quad
\bar\omega(t)=t^{p-\frac{1}{2}}e^{-\frac{at^2}{2}+bt}.
\label{6.4}
\end{eqnarray}
Note now that the relation (\ref{4.1}) for wavefunctions
implies the analogous relation for the corresponding polynomials
\begin{eqnarray}
P_m(t)\sim \bar P_{M-m}(-t)
\label{6.5}
\end{eqnarray}
whose comparison with (\ref{6.3}) gives
\begin{eqnarray}
\int_{0}^{\infty}P_m(t)P_n(t)\omega(t)dt=0,\quad
\int_{-\infty}^{0}P_m(t)P_n(t)\omega(t)dt=0,\quad
m\neq n.
\label{6.6}
\end{eqnarray}
So, we see that polynomials $P_m(x),\
m=0,1,\ldots,[M/2]$ form an orthogonal system with the
weight function $\omega(x)$ on two different intervals $[-\infty,0]$
and $[0,\infty]$. 

Note that the existence of two intervals on which the
polynomials $P_n(x), \ n=0,\ldots, [M/2]$ form an
orthogonal system is closely related to the facts that all
these polynomials are of the same degree $[M/2]$ and their
number $[M/2]+1$ is finite. Indeed, in this case, 
the total number of coefficients
of these polynomials (excluding the leading ones which can
be considered as normalization factors) is $[M/2]([M/2]+1)$.
Requiring the orthogonality of polynomials on a certain
interval, we obtain $[M/2]([M/2]+1)/2$ equations for their coefficients.
In this case the number of equations is only one half of
the total number of unknowns which prevents one from
determining these polynomials uniquely. However, if we require
the orthogonality of these polynomials on two different intervals,
then the number of equations will exactly coincide with the
number of unknowns, and the polynomials can be determined uniquely.
Note that this property is typical for all non-standard orthogonal 
polynomials associated with quasi-exactly solvable models from ref.
\cite{ush}.

\section{Conclusion}

The duality property of the simplest quasi-exactly solvable
model (\ref{2.1}) and its applications discussed in
sections 3 -- 6 can easily be extended to 
all one-dimensional quasi-exactly solvable models listed in
ref. \cite{ush}. In order to demonstrate this, remember that these
models are associated with the class
of equations
\begin{eqnarray}
\omega\left\{(t+\bar a)t(t-a)
\frac{\partial^2}{\partial t^2}+ 
(At^2+Bt +C)
\frac{\partial}{\partial t}-
M(M+A-1)\right\}P(t)=EP(t)
\label{7.1}
\end{eqnarray}
(or their degenerate forms) having $M+1$ polynomial solutions for any $M$. 
The transition of equation (\ref{7.1}) to the Schr\"odinger 
form can be performed after introducing new variables $x$
and new functions $\Psi(x)$ by the formulas
\begin{eqnarray}
x=\int\frac{dx}
{\sqrt{-\omega(t+\bar a)t (t-a)}}.
\label{7.3}
\end{eqnarray}
and
\begin{eqnarray}
\Psi(x)=P(t)\left(\frac{dx}{dt}\right)^{\frac{1}{2}}
\exp\left\{\int\frac{At^2+Bt+C}{\omega(t+\bar a)t(t-a)}dt\right\}
\label{7.2}
\end{eqnarray}
in which the variable $t$ belongs to the
interval $[0,a]$ (for $\omega>0$) or to the interval $[a,\infty]$
(for $\omega<0$). The set of parameters $\omega, a, \bar a, A, B, C$
determines the form of the resulting potential. 
From (\ref{7.1}) it immediately follows that the transformation
$t\rightarrow -t$, $E\rightarrow -E$  reduces the quasi-exactly
solvable model characterized by the set of parameters 
$\{\omega, a, \bar a, A, B, C\}$ to another model
characterized by the set $\{\omega, \bar a,  a, A, -B, C\}$.
This transformation is nothing else than the generalization
of the duality transformation for model (\ref{2.1}).

It would be an interesting problem to investigate Lie algebraic
origin of the duality property of quasi-exactly solvable models
and find a multi-dimensional generalization of this phenomenon.
We intend to consider these questions in fortcoming publications.

\section{Acknowledgements}

The authors are grateful to K. Smoli\'nski for valuable remarks.

\end{document}